\documentclass[%
reprint,
superscriptaddress,
nofootinbib,
amsmath,amssymb,
aps,
]{revtex4-2}
\usepackage{graphicx}
\usepackage{dcolumn}
\usepackage{bm}
\usepackage[hidelinks]{hyperref}
\hypersetup{
colorlinks = true,
linkcolor = blue,
anchorcolor = blue,
citecolor = blue,
urlcolor = blue
}

\usepackage{lineno}
\usepackage{amssymb}
\usepackage{amsmath}
\usepackage{epstopdf}
\usepackage{natbib}
\usepackage{xcolor}
\usepackage{natbib}
\usepackage{diagbox}
\usepackage{multirow}
\usepackage{threeparttable}
\newcommand{\apjl}{Astrophys. J. Lett.}
\newcommand{\mnras}{Monthly Notices of the Royal Astronomical Society}
\newcommand{\araa}{Annual Review of Astronomy and Astrophysics}
\newcommand{\jcap}{Journal of Cosmology and Astroparticle Physics}
\newcommand{\aap}{Astronomy \& Astrophysics}
\newcommand{\baas}{Bulletin of the American Astronomical Society}
\begin{document}

	\preprint{APS/123-QED}
	
	\title{Constraining the Charge of a Black Hole with Electromagnetic Radiation from a Black Hole-Neutron Star System}
	
	\author{Hao-Yu Yuan}
	\affiliation{Guangxi Key Laboratory for Relativistic Astrophysics, School of Physical Science and
		Technology, Guangxi University, Nanning 530004, China}
    \affiliation{Department of Astronomy, School of Physics, Huazhong University of Science and Technology, Wuhan, 430074, China}
	
	\author{Hou-Jun L\"{u}}
	\email{lhj@gxu.edu.cn}
	\affiliation{Guangxi Key Laboratory for Relativistic Astrophysics, School of Physical Science and
		Technology, Guangxi University, Nanning 530004, China}
	
	\author{Jared Rice}
	\affiliation{Department of Mathematics and Physical Science, Southwestern Adventist
University, Keene, TX 76059, USA}
	
	\author{En-Wei Liang}
	\affiliation{Guangxi Key Laboratory for Relativistic Astrophysics, School of Physical Science and
		Technology, Guangxi University, Nanning 530004, China}
	
	\date{\today}
	
\begin{abstract}
Black hole-neutron star (BH-NS) mergers are expected to emit gravitational-wave (GW) and electromagnetic (EM) counterparts when the NS is tidally disrupted or plunges into the BH. Recently, GW 200105 and GW200115 were claimed as originating in BH-NS mergers, even GW 200105 remains in debate. Several optical source candidates are reported to possible associate with the two GW events, but not confirmed yet. In this work, we assume that the BH is charged (the NS is naturally charged) and try to constrain the charge of the BH by using the possible associated EM emission from the charged BH and NS system working in the inspiral regime. We adopt electric and magnetic dipole radiations for the binaries which power a Poynting-flux-dominated outflow to accelerate electrons. Then, it produces the observed EM radiation via synchrotron radiation. We find that the conversion efficiency in the X-ray band is much higher than that of the ultraviolet (UV), near-infrared, and radio bands. The estimated maximum charge-to-mass ratio (the charge for unit mass) of the BH is $1.12\times 10^{-6}$ and $1.53\times 10^{-6}$ esu for the binary systems of GW200105 and GW200115, respectively, if magnetic field strength $B_{p}\lesssim ~10^{16}$ G and period $P>~1$ ms for the NS spin.
\end{abstract}	
\maketitle
	
\section{introduction}
Compact binary coalescences, including black hole-black hole (BH-BH), black hole-neutron star(BH-NS), and neutron star-neutron star (NS-NS) mergers, are expected to be a strong source for the production of gravitational wave (GW) radiation in the Universe \citep[][for a review]{Berger(2014)}. Catching the GW signal from such binary systems is the main target of current ground-based GW detectors, e.g., advanced Laser Interferometer Gravitational-Wave Observatory \citep[LIGO;][]{LIGO Scientific Collaboration et al.(2015)}, advanced Virgo Interferometer \citep[Virgo;][]{Acernese et al.(2015)}, and KAGRA \citep{Aso et al.(2013)}. The first detected GW signals from the two events GW 150914 and GW 151226 with LIGO are proposed to be black hole binary mergers \citep{Abbott et al.(2016)}. More interestingly, weak electromagnetic (EM) transients associated with the two GW events in the $\gamma$-ray band were claimed to be detected, but this is still highly debated \citep{Connaughton et al.(2016), Lyutikov(2016), Greiner et al.(2016)}. On 17 August 2017, advanced LIGO and Virgo first directly detected both the GW signal (GW170817) and its EM counterparts (e.g., GRB 170817A and kilonova AT2017gfo) from the proposed merger of a binary NS system \citep{Abbott et al.(2017a),Abbott et al.(2017b), Goldstein et al.(2017), Savchenko et al.(2017), Chornock et al.(2017), Coulter et al.(2017), Covino et al.(2017), Zhang et al.(2018)}.

Despite the growing number of detected GW events, only several candidate BH-NS mergers have been reported, namely GW190426 \citep{Li et al.(2020)}, GW190814 \citep{Abbott et al.(2020)}, GW190917 \citep{The LIGO Scientific Collaboration et al.(2021a)}, GW191219 \citep{Gompertz et al.(2022)}, and GW200210 \citep{Kruckow Han(2021)}. Recently, GW 200105 and GW200115 were claimed as originating from the merger of BH-NS systems by the LIGO-Virgo-KAGRA (LVK) Collaboration during the third observing run (O3) stage \citep{Abbott et al.(2021a)}, even GW 200105 remains in debate. Unfortunately, no conclusive EM counterparts have been detected yet, except for several possible optical source candidates associated with the two GW events \citep{Anand et al.(2021), Alexander et al.(2021), Coughlin et al.(2020), Gompertz et al.(2020), Kasliwal et al.(2020), Page et al.(2020), Thakur et al.(2020), Dobie et al.(2022)}.

Whether or not EM counterparts accompany BH-NS mergers remains an open question. From the theoretical point of view, NSs can be tidal disrupted if the tidal disruption radius of the NS is larger than the innermost stable circular orbit (ISCO) radius of the BH and the massive debris can be expelled or accreted onto the newborn BH to power the EM counterparts \citep{Foucart(2012), Foucart et al.(2013)}. However, the results of numerical simulations have shown that the conditions for NS disruption within inspiraling BH-NS binaries are extremely strict. For example the NS equations of state (EOS) need to be stiff enough or the BH spin projected to the orbital angular momentum needs to be extremely large \citep{Fragione(2021), DOrazio et al.(2021)}. On the other hand, the observed GW and EM data from LIGO, Virgo, and other telescopes are not likely to support the stiff NS EOS and the extremely high BH spin \citep{Lasky(2014), Lyu(2015), Lyu(2017), Abbott et al.(2018), Abbott et al.(2021b)}. Given the observational evidences, NSs plunging into the BHs during BH-NS mergers seems to be a natural physical processes during stellar evolution.

Within in this scenario, reference \citep{Zhang(2019)} proposed that EM radiation associated with the GW signal of the BH-NS merger can be produced if at least one of the members of BH-NS merger is charged. The system can raise intense electric dipole and magnetic dipole radiation near the orbital plane, and it propagates outward with a Poynting-flux-dominated outflow \citep{Zhang(2019)}. Finally, magnetic energy is dissipated to accelerate the electrons and can power the EM radiation with different radiation mechanisms (e.g., synchrotron radiation). On the other hand, the physical properties of a BH can be described simply with three parameters (mass, angular momentum, and charge). The mass and angular momentum of the BH can be roughly measured
based on abundant observational data, but the charge of a BH is poorly understood \citep{Shapiro Teukolsky(1983), Inayoshi et al.(2020), Feng et al.(2022)}. Motivated by reference \citep{Zhang(2019)}, one interesting question is whether or not we can constrain the charge of the BH using either the observed EM radiation or radiation upper limits in the BH-NS binary system if we assume that the BH is charged.

In this paper, we try to constrain the charge of BHs using the observed upper limits of EM radiation for GW200105 and GW200115 by assuming the BHs in the BH-NS systems are charged. We organize this paper as follows. The theoretical derivation of EM counterpart production during the merger of a NS with a charged BH is presented in Section \ref{Sec: EM-counterpart}. In section \ref{Sec: GW and EM}, we describe the observations of GW200105 and GW200115. The constrained results are shown in Section \ref{Sec: results}. The conclusions are drawn in Section \ref{sec:Conclusion} with some additional discussions. Throughout this paper, we use cgs unit and adopt a concordance cosmology with parameters $H_{0} = 67.4~\rm ~km~s^{-1}~ Mpc^{-1}, ~\Omega_{M} = 0.315, ~and ~\Omega_{\Lambda} = 0.685$ \citep{Planck Collaboration et al.(2020)}.

\section{General theory of charged BH-NS mergers}\label{Sec: EM-counterpart}
\subsection{Electric dipole and magnetic dipole radiations}
Two different channels for the formation of a BH-NS system are discussed in literatures \citep{Kyutoku et al.(2021), Shibata Hotokezaka(2019),Benacquista Downing(2013)}, but they are remaining in debate. One is binary systems with each member of the binary undergoing massive star collapse when the thermal force of the star can not support its gravitational force \citep[for a review][]{Kyutoku et al.(2021), Shibata Hotokezaka(2019)}. The other one is the dynamical interactions between BHs and NSs in globular clusters which contain many compact stars (e.g., BH and NS). Then, these interactions can produce the binary systems that contain two or more compact objects \citep{Benacquista Downing(2013)}.

After the BH-NS binary is formed, the orbital separation between the objects gradually decreases until coalescence because of the energy loss due to GW emission. The NS has a strong surface magnetic field and a rapid rotation, and its surface magnetic field changes with time as it is orbiting with the BH. The changing NS magnetic field can produce an electric field, so that both a magnetic field and an electric field live on the surface of the NS. For our BH-NS system, we assume that NS plunges into the BH without tidal disruption and that the BH is charged. If this is the case, the EM signals may be produced via both electric dipole and magnetic dipole radiation and propagate
outward with a Poynting-flux-dominated outflow.

Following the method of references \citep{Zhang(2016)} and \citep{Zhang(2019)}, $M_{i},~\hat{q}_{i}$ are defined as the mass and relative charge of the BH or NS, respectively. Here, the subscript $i$ represents either the BH or the NS. $\hat{q}_{i}$ is defined as the ratio $Q_{i}/M_{i}$ \citep{Zhang(2016)}, where $Q_{i}$ and $M_{i}$ are the absolute charge and mass of the BH or NS, and $G$ is the gravitational constant. Moreover, we also define other parameters as follows:\\ \centerline{Total mass: $M=M_{BH}+M_{NS}$;}\\
\centerline{Mass ratio: $q=M_{BH}/M_{NS}$;}\\
\centerline{Reduced mass: $M_{r}=M_{BH}M_{NS}/(M_{BH}+M_{NS})$;}\\
\centerline{Chirp mass: $M_{c}=M_{r}^{3/5}M^{2/5}$;}\\
\centerline{Horizon mass: $M_{h}=M_{r}^{2/5}M^{3/5}$.}

The charged BH-NS system can give rise to electric dipole and magnetic dipole radiation due to the interaction between the magnetosphere of the binaries during inspiral. We present more details of both electric dipole and magnetic dipole radiation in the following.

{\bf (1) Luminosity of electric dipole radiation:}

In general, considering that just one member of a BH-NS binary is charged (e,g., NS is charged), the electric dipole radiation luminosity can be written as by Larmor formula\citep{Zhang(2016),Deng et al.(2018),Zhang(2019)}
\begin{equation}
\begin{split}
\label{eq_1}
L_{e,NS}&=\frac{2Q_{NS}^{2}\lvert \ddot{r}_{NS} \rvert^{2}}{3c^{3}}\\
&=\frac{1}{24}\frac{c^{5}}{G^2}\hat{q}_{NS}^{2}\left[\frac{r_{s}(M_{BH})}{a}\right]^{2}
\left[\frac{r_{s}(M_{NS})}{a}\right]^{2}
\end{split}
\end{equation}
where $c$ is speed of light, and $r_{s}(M_{i})=\frac{2GM_{i}}{c^{2}}$. Hence, $r_{s}(M_{BH})$ and $r_{s}(M_{NS})$ are the Schwarzschild radii of the masses $M_{BH}$ and $M_{NS}$, respectively. $\lvert \ddot{r}_{NS} \rvert=\frac{GM_{BH}}{a^{2}}$ is the amplitude of the acceleration of the NS and $a$ is the separation of the BH-NS binary. Similarly, one may easily write the luminosity if only the BH is charged.  However, if both NS and BH are charged, one should consider not only the contributions from charged NS and charged BH itself, but also the cross-term effect between the charged BH and charged NS. The total dipole moment of the system is $d=Q_{\rm NS}r_{\rm NS}+Q_{\rm BH}r_{\rm BH}$ by choosing the origin of coordinates at the centre of mass, where $r_{\rm NS}$ and $r_{\rm BH}$ are the radius vector of neutron star and black hole, respectively. Based to Larmor formula, one has
\begin{equation}
\begin{split}
\label{eq_Lar}
L_{e}&=\frac{4}{3c^3}\ddot{d}^2
\end{split}
\end{equation}
\begin{equation}
\begin{split}
\label{eq_ddot_d}
M_i\ddot{r}_i=\pm\frac{GM_{NS}M_{BH}}{r^2}\mp\frac{Q_{NS}Q_{BH}}{r^2},
\end{split}
\end{equation}
where $r=r_{NS}-r_{BH}$ is the relative position of them. We can derive the luminosity of electric dipole radiation for both charged BH and charged NS by adopting $a\approx |r|$ \citep{Cardoso et al.(2016), Liu et al.(2020), Christiansen et al.(2021)},
\begin{equation}
\begin{split}
\label{eq_L_e}
L_{e}
&\approx\frac{4}{3}\frac{c^{5}}{G^{2}}(\hat{q}_{NS}
-\hat{q}_{BH})^2\left(1-\frac{\hat{q}_{NS}\hat{q}_{BH}}{G}\right)^2\\
&\left[\frac{r_{s}(M_{BH})}{2a}\right]^{2}
\left[\frac{r_{s}(M_{NS})}{2a}\right]^{2}.
\end{split}
\end{equation}

Here, we ignore the effect of EM force on the orbital evolution, and only consider the gravitational force between the BH and NS to calculate the orbital evolution of the systems, because the charge of BH is small enough. We compare with the luminosity between EM and GW at different radius for different charge, and derive the critical condition for the validity of this assumption. The GW luminosity is expressed as,
\begin{equation}
\begin{split}
\label{eq_L_GW}
L_{GW}
=\frac{32}{5}\frac{G^{4}}{c^{5}}\frac{M_r^2 M^3}{a^5}f(e).
\end{split}
\end{equation}
where $f(e)=(1+\frac{73}{24}e^{2}+\frac{37}{96}e^{4})/(1-e^{2})^{7/2}$ is a coefficient for elliptic orbits, and $e$ is the orbital eccentricity. We adopt a reasonable value of $e=0$ in the calculations below\footnote{The orbital eccentricity is decreasing during inspiral, and it would be close to zero due to GW or EM radiations, and ejected energy (the electric dipole and magnetic dipole radiations) are close to maximum values with the increasing of orbital frequency when the orbital eccentricity is close to zero. Thus, the ejected energy at this stage can represent the total energy which ejected during the inspiral and coalescence. Combining these accounts, orbital eccentricity equal to zero is a reasonable approximation \citep{Zhang(2019)}.}\citep{Zhang(2019)}.

In addition, the orbital separation $a$ gradually decreases due to the energy loss of GW
emission during inspiral, and the rate of change of $a$ can be written as,
\begin{equation}
\begin{split}
\label{eq_dota}
\frac{da}{dt}&=-\frac{64}{5}\frac{G^{3}}{c^{5}}\frac{M_{NS}\cdot
M_{BH}(M_{NS}+M_{BH})}{a^{3}}f(e)\\
&=-\frac{64}{5}\frac{G^{3}}{c^{5}}\frac{M_{r}M^{2}}{a^{3}}f(e)
\end{split}
\end{equation}
\\

{\bf (2) Luminosity of magnetic dipole radiation:}

When a charged BH-NS binary inspiral, a loop current can be produced because of the interaction between their magnetospheres. Generally speaking, BH is considered to be electrically neutral. However, if the BH is in the non-electricity neutral environment or formed in the strong magnetic fields, the BH would be charged and generated magnetospheres \citep{Cardoso et al.(2016), Zajacek et al.(2019)}. In classical electromagnetism, the magnetic dipole moment is the product of the current and effective area of the loop \citep{Zhang(2016)},
\begin{equation}
\begin{split}
\label{eq_mu}
\mu&=\frac{\pi}{c}I\left(\frac{a}{2}\right)^{2}\\
&=\sqrt{GMa}~\frac{Q_{BH}+Q_{NS}}{8c},
\end{split}
\end{equation}
where $I=(Q_{BH}+Q_{NS})/P_s$ is the loop current, and $P_s=2\pi a^{3/2}/\sqrt{GM}$ is the orbital
period\citep{Bildsten Cutler(1992),Kochanek(1992)}. The luminosity of magnetic dipole radiation can be expressed as
\begin{equation}
\begin{split}
\label{eq_LB}
L_{m}&=\frac{2\ddot{\mu}^{2}}{3c^{3}}\\
&\approx8.56\times 10^{2}\frac{G^{13}M_{r}^{4}M^{9}}{c^{25}a^{15}}(\hat{q}_{BH}\cdot
M_{BH}+\hat{q}_{NS}\cdot
M_{NS})^2
\end{split}
\end{equation}
where $\ddot{\mu}$ is the second derivative of magnetic dipole moment.

\subsection{Calculation of the charge of a NS}
Reference \citep{Goldreich Julian(1969)} proposed the distribution of spatial charge density ($\rho_{e}$) of a magnetized NS, $\rho_{e}=-\mathbf{\Omega} \cdot \mathbf{B}/(2\pi c)$, where $\mathbf{\Omega}$ and $\mathbf{B}$ are the angular velocity and magnetic field strength of the NS, respectively. Assuming that the surface magnetic field of the NS is a dipole field, $B=B_pR_{NS}^3/(r^3)(3cos^2\theta +1)^{1/2}$, we adopt a simple situation that $\mathbf{\Omega}\cdot \mathbf{B}<0$. The charge of the NS ($Q_{NS}$) consists of two parts: the charge within the magnetosphere ($Q_{mag}$) and inside the NS ($Q_{in}$). For $Q_{mag}$, one needs to integrate along the distance from the NS surface ($R_{NS}$) to $r$, which is the distance between the reference point and the NS center,
\textbf{\begin{equation}
\begin{split}
\label{eq_Qmag}
Q_{mag}&=\int^{2\pi}_{0}\int^{\pi}_{0}\int^{r}_{R_{NS}}-\frac{\mathbf{\Omega} \cdot
\mathbf{B}}{2\pi c}r^2sin\theta drd\varphi d\theta\\
&=\frac{\Omega B_pR_{NS}^3}{c}\int^{r}_{R_{NS}}\frac{1}{r}dr\int^{\pi}_{0}cos\theta
sin\theta\sqrt{3cos^2\theta +1}d\theta,
\end{split}
\end{equation}}
where $B_{p}$ is the strength of surface dipole magnetic field of the NS, and $\theta$ is azimuthal angle. Because the antisymmetry of the integrand around $\pi/2$ for the integral over $\theta$, one can guarantee that the value of integrating for $\theta$ is zero, namely, $Q_{mag}=0$. Similarly, if the NS is uniformly magnetized, one can calculate $Q_{in}$ as \citep{Ruderman Sutherland(1975)},
\begin{equation}
\begin{split}
\label{eq_Qin}
Q_{in}=\frac{2}{3}\Omega B_{p} R_{NS}^{3}.
\end{split}
\end{equation}
The total charge of the NS depends on the properties of the NS (e.g., angular velocity, dipole magnetic field strength, and the radius), and can be written as
\begin{equation}
\begin{split}
\label{eq_QNS}
Q_{NS}=Q_{mag}+Q_{in}.
\end{split}
\end{equation}

\subsection{Synchrotron radiation of Poynting-flux-dominated outflow}
In this section, we will discuss how the BH-NS system will radiate energy. For either magnetic dipole or electric dipole radiation, the radiation frequency is equal to the orbital frequency of $\sim {\rm kHz}$. However, the orbital frequency is much lower than the intrinsic frequency of interstellar medium \citep[$\sim 10^{4}{\rm~Hz}$,][]{Cardoso et al.(2021)}. Here, we do not consider the affect of the high interstellar medium ($\sim 1\rm~cm^{-3}$). The intrinsic (or oscillation) frequency of interstellar medium ($\sim 10^4 n_e^{1/2}\rm~Hz$) originates from the thermal motion of electrons within the Debye length, where $n_e$ is the electron number density per $\rm cm^3$ in the interstellar medium. The low-frequency electromagnetic radiation can be prevented from propagating when EM frequency is less than the intrinsic frequency of interstellar medium. Therefore, it is inevitable that the radiation would be trapped in the vicinity of the binary. In practice, the energy of the dipole radiation can propagate outward with a Poynting-flux-dominated outflow. The Poynting luminosity at a radius ($r_p$) is written as \citep{Beniamini Giannios(2017), Giannios Spruit(2005)},
\begin{equation}
\begin{split}
\label{eq_Lp}
L_{p}&=c\frac{(r_p B_{jet})^{2}}{4\pi}\\
&=(L_{e}+L_{m})\left(1-\frac{\Gamma}{\Gamma_{sat}}\right)\\
&=L_{tot}\left(1-\frac{\Gamma}{\Gamma_{sat}}\right),
\end{split}
\end{equation}
where $B_{jet}$ and $\Gamma$ are the jet magnetic field strength and bulk Lorentz factor within the outflow, respectively.

The total energy from the electric dipole and magnetic dipole radiation ($L_{tot}=L_{e}+L_{m}$) is used to inject the outflow, and $\Gamma_{sat}$ is the bulk Lorentz factor at the saturation radius $r_{sat}$ of the outflow. Subsequently, the magnetic energy will be dissipated gradually via magnetic reconnection to accelerate electrons. Moreover, particle-in-cell (PIC) simulations suggest that the energy spectrum of accelerated electrons is roughly a power-law
distribution \citep{Sironi Spitkovsky(2014), Guo et al.(2015), Werner et al.(2016)},
\begin{equation}
\begin{split}
\label{eq_dNe}
N(\gamma_{e})d\gamma_{e}\propto \gamma^{-p}_{e}d\gamma_{e}, \gamma_{e}\geq\gamma_{m},
\end{split}
\end{equation}
where $\gamma_{e}$ and $\gamma_{m}$ are the Lorentz factor and minimum Lorentz factor of
accelerated electrons, respectively, and $p$ is the power-law index of accelerated electrons. Here, we adopt $p=4\sigma^{-0.3}$ according to a reasonable fit for the results of numerical calculations, and $\sigma$ is the magnetization parameter of the outflow. Qualitatively, it is the ratio between magnetic energy and kinetic energy in the outflow, namely $\sigma=L_{p}/L_{k}$. As the energy of accelerated electrons is dissipated, it can produce broadband radiation via different radiation mechanisms (e.g., synchrotron radiation and inverse compton scattering). In this paper, we consider only the pure synchrotron radiation\footnote{The synchrotron radiation is the main radiation mechanism in GRB study via the observational data \citep{ZhangB(2018)}. Here, we do not consider other radiation mechanisms, such as curvature radiation, compton scattering, and inverse Compton scattering.} which can generate multi-band afterglow of GRBs \citep[from X-ray to radio,][]{Xiao Dai(2017), Beniamini Giannios(2017)}.

Reference \citep{Sari et al.(1998)} considered a relativistic shock propagating through a uniform cold medium, and assumed that the shock undergoes adiabatic and radiative hydrodynamic evolution. Within this scenario, they calculated the radiated spectrum for two different regions, e.g., fast cooling and slow cooling cases. In order to distinguish those two spectral regions, one define a parameter $\gamma_c$ which is the critical Lorentz factor for electron synchrotron radiation. The energy loss through synchrotron radiation becomes significance when the minimum Lorentz factor of electrons exceed the threshold exceed. So that, one can separates those two regions by comparing with $\gamma_c$, e.g., $\gamma_m > \gamma_c$ (fast cooling) and $\gamma_m < \gamma_c$ (slow cooling). We list the luminosity ($L_{\nu}$) at each frequency ($\nu$) as follows for these two cases.
\begin{enumerate}
  \item {\bf Fast cooling regime:}
\begin{equation}
\begin{split}
\label{eq_Lnufast}
L_{\nu}=\left\{
\begin{aligned}
&L_{\nu,max}\left(\frac{\nu}{\nu_{c}}\right)^{\frac{1}{3}}, & \nu<\nu_{c}\\
&L_{\nu,max}\left(\frac{\nu}{\nu_{c}}\right)^{-\frac{1}{2}}, & \nu_{c}<\nu<\nu_{m}\\
&L_{\nu,max}\left(\frac{\nu_{m}}{\nu_{c}}\right)^{-\frac{1}{2}}\left(\frac{\nu}{\nu_{m}}\right)^{-\frac{p}{2}},
& \nu_{m}<\nu<\nu_{max}
\end{aligned}
\right.
\end{split}
\end{equation}

  \item {\bf Slow cooling regime:}
\begin{equation}
\begin{split}
\label{eq_Lnuslow}
L_{\nu}=\left\{
\begin{aligned}
&L_{\nu,max}\left(\frac{\nu}{\nu_{m}}\right)^{\frac{1}{3}}, & \nu<\nu_{m}\\
&L_{\nu,max}\left(\frac{\nu}{\nu_{m}}\right)^{-\frac{p-1}{2}}, & \nu_{m}<\nu<\nu_{c}\\
&L_{\nu,max}\left(\frac{\nu_{c}}{\nu_{m}}\right)^{-\frac{p-1}{2}}\left(\frac{\nu}{\nu_{c}}\right)^{-\frac{p}{2}},
& \nu_{c}<\nu<\nu_{max}
\end{aligned}
\right.
\end{split}
\end{equation}
\end{enumerate}
where $L_{\nu,max},~\nu_{m},~\nu_{c}$ and $\nu_{max}$ correspond to the maximal luminosity, typical frequency, cooling frequency, and the maximal frequency, respectively. The maximal synchrotron spectral luminosity,
\begin{equation}
\begin{split}
L_{\nu,max}&=\frac{m_ec^2\sigma_T B_{jet} N_e}{3e}\\
&=\frac{m_ec^2\sigma_T N_e}{3e}\left[\frac{4\pi L_{tot}}{cr_p^2}\left(1-\frac{\Gamma}{\Gamma_{sat}}\right)\right]^{1/2}
\end{split}
\end{equation}
where $\sigma_T$, $e$ and $N_e$ are the Thomson scattering cross-section, the electron charge, and the total number of emitting electrons in the jet at $r_p$, respectively. Eq. (\ref{eq_Lnufast}) and (\ref{eq_Lnuslow}) are invalid as long as the frequency is below the synchrotron self-absorption
(SSA) frequency $\nu_{a}$, and
\begin{equation}
\begin{split}
\label{eq_nua}
\frac{2\nu^{2}_{a}}{c^{2}}\gamma_{a}\Gamma m_{e}c^{2}\frac{\pi
r_{p}^{2}}{\Gamma^{2}}=L_{\nu_{a}},
\end{split}
\end{equation}
where $m_{e}$ and $\gamma_{a}$ are the electron mass and Lorentz factor corresponding to $\nu_{a}$, respectively. The SSA effect might play an important role when $\nu < \nu_{a}$, and the spectral shape become $L_{\nu}\propto \nu^{11/8}$ \citep{Granot Sari(2002)}.

\section{Observations of the GW200105 and GW200115 events}\label{Sec: GW and EM}
During the third observing run (O3), the LVK reported that two GW events (GW200115 and GW200105) are originated from BH-NS binaries \citep{Abbott et al.(2021a)}, but the GW 200105 remains in debate later due to the high likelihood of detector noise \citep{The LIGO Scientific Collaboration et al.(2021b)}. Based on the GW signals from the two events, the inferred mass of their primary are $8.9^{+1.2}_{-1.5}M_{\odot}$ and $5.7^{+1.8}_{-2.1}M_{\odot}$, respectively. These primary masses are well in excess of the maximum NS mass, but fall into the mass range of BHs. On the other hand, the mass of the companions are $1.9^{+0.3}_{-0.2}M_{\odot}$ and $1.5^{+0.7}_{-0.3}M_{\odot}$, respectively. These are consistent with the mass range of known NSs and are below the maximum NS mass. Moreover, the LVK Collaboration claims that GW observations of LIGO and Virgo have led to the identification of five BH-NS candidates, GW190426 \citep{Li et al.(2020)}, GW190814 \citep{Abbott et al.(2020)}, GW190917 \citep{The LIGO Scientific Collaboration et al.(2021a)}, GW191219 \citep{Gompertz et al.(2022)}, and GW200210 \citep{Kruckow Han(2021)}. The inferred primary mass, secondary mass, chirp mass, mass ratio, and redshift are collected in Table \ref{tab:GW properties} for these GW events and their candidate BH-NS merger objects.

Searching for the EM counterparts of a BH-NS merger has so far remained main targets and expectations of space and ground telescopes. There are no EM counterparts associated with the GW candidates of BH-NS merger that were claimed by the LVK Collaboration. For the GW200105 and GW200115 events, no significant signals were caught in the $\gamma-$ray band above the background because of the short-duration and rapid decline of the expected short GRB\footnote{The other possibility is that no gamma-ray signals were produced when they are in non-disrupting systems.}. \citep{Doyle et al.(2020), Veres et al.(2020), Lucarelli Agile Team(2020), Barthelmy Swift Team(2020), HAWC Collaboration(2020a), HAWC Collaboration(2020b), Shenoy AstroSat CZTI Collaboration(2020), Yamaoka et al.(2020), Ridnaia et al.(2020a), Ridnaia et al.(2020b), Ursi et al.(2020a), Ursi et al.(2020b), Ferrigno et al.(2020), Goldstein et al.(2020), Crnogorcevic Fermi-LAT Collaboration(2020), Sakamoto Swift Team(2020), Shimizu et al.(2020), Sugita et al.(2020), Kawai MAXI Team(2020)}. However, the Swift/XRT team claimed that several possible X-ray emission candidates within the localization area of the GW trigger S200115j were caught \citep{Evans Swift Team(2020a), Evans Swift Team(2020b)}. Afterwards, possible associated optical/IR observations have been reported within an error box larger than that of the error boxes of GW200105 and GW200115, though they cannot be confirmed.

Global MASTER-Net telescopes started to scan the GW200105 error box  $\sim3.2$ hours after the GW trigger, and got a series of upper limits in V- and C-band\footnote{C-band is a clear (unfiltered) band.} \citep{Lipunov et al.(2020)}. The Zwicky Transient Facility (ZTF) inspected the localization area that covered $\sim51.7\%$ of the enclosed area of the GW trigger S200105ae in the g- and r-band, and reported more than 20 optical transients as the possible EM candidates of GW200105 \citep{Stein et al.(2020), Ahumada(2020)}, although several other observations stated that there
are five candidates that actually are consistent with supernovae (SNe) \citep{Valeev Font(2020), Castro-Tirado Font(2020), Andreoni(2020)}. For GW200115, the ZTF and Global Relay of Observatories Watching Transients Happen (GROWTH) collaborations accidentally observed the localization region of the GW trigger S200115j and covered 22\% of the localization probability \citep{Anand et al.(2020)}. \citet{Srivastav Smartt(2020)} also reported the transients within one of the eastern lobes of this trigger with the Pan-STARRS2 telescope. The optical/IR observations reported by GCN are summarized in Table \ref{tab:EM candidates 200105} and \ref{tab:EM candidates 200115}.

\section{Results}\label{Sec: results}
To date, there is indeed not directly evidence to observe the EM counterparts of BH-NS
systems, only report the optical candidates or upper limits of EM counterparts in the two events GW200105 and GW200115. If this is the case, one can constrain the BH charge according the hypothesis of charged BH-NS systems by adopting the upper limits. In this section, we assume that either the observed optical transients above or even upper limits as the EM counterparts of GW200105 and GW200115 events, and adopt those observed EM counterparts to constrain the BH charge by given the hypothesis of a charged BH-NS system.

\subsection{Conversion efficiency of electric dipole and magnetic dipole radiation}
\label{subsec: Efficiency of Conversion} The charged BH and NS system can produce both electric dipole and magnetic dipole radiation due to the interaction between the magnetospheres of the objects during inspiral. The total energy from the electric dipole and magnetic dipole radiation ($L_{tot}$) is used to inject into the outflow, and it propagates outward with a Poynting-flux-dominated outflow. Within the synchrotron radiation scenario, the conversion efficiency ($\eta$) is defined as the ratio between the luminosity of the observed energy-bands and total luminosity (e.g., the sum of electric and magnetic dipole radiation),
\begin{equation}
\begin{split}
\label{eq_eta}
\eta=\frac{\int_{\nu_{1}}^{\nu_{2}}L_{\nu}d\nu}{L_{tot}},
\end{split}
\end{equation}
where $\nu_{1}$ and $\nu_{2}$ are the frequency range of a certain energy-band or detector, respectively.

\begin{figure*}
\centering
\includegraphics [angle=0,scale=0.6] {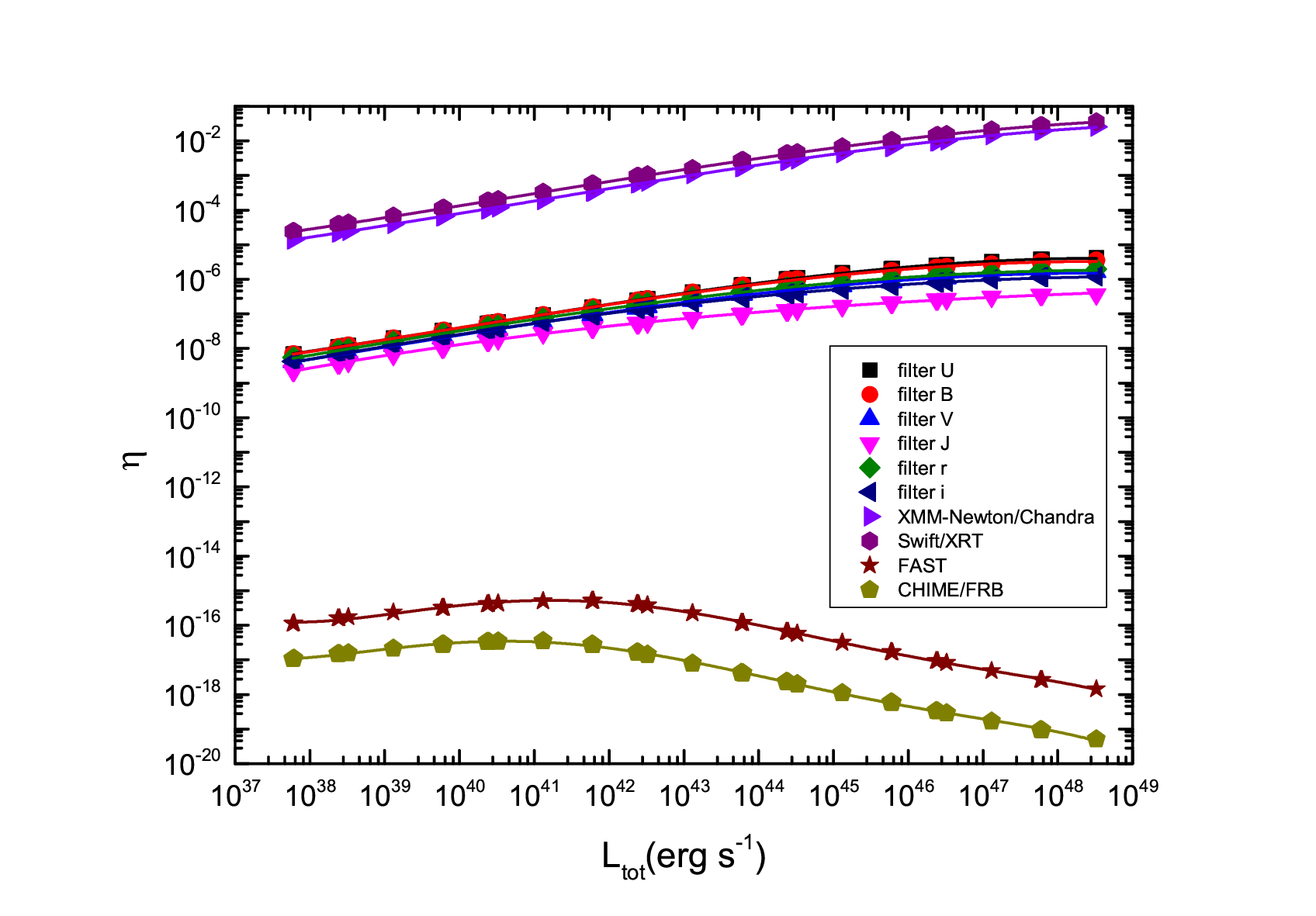}
\caption{$\eta$ as a function with $L_{tot}$ for different energy bands.}
\label{fig:L_eta}
\end{figure*}

The conversion efficiency is dependent on the frequency and the synchrotron radiation luminosity $L_{\nu}$. For a given energy band, Eq.(\ref{eq_eta}) can be calculated by adopting Eqs.(\ref{eq_Lnufast}), (\ref{eq_Lnuslow}), and (\ref{eq_nua}). Here, we adopt the $\Gamma_{\rm sat}=\sigma_0 \Gamma_0=\sigma^{3/2}_{0}=1000$ to calculate the $L_{\nu, max}$, where $\Gamma_0$ and $\sigma_0$ are the initial Lorentz factor and magnetization parameter of the outflow, respectively. One can calculate the conversion efficiency as a function of the given total luminosity \citep{Xiao Dai(2017)}. Fig.\ref{fig:L_eta} shows the conversion efficiency as function of $L_{tot}$ from the X-ray to radio bands. It is clear to see that the conversion efficiency in the X-ray band is much higher than that of the ultraviolet (UV), near-infrared, and radio bands, and increases with $L_{tot}$ in X-ray, UV, and near-infrared. However, the conversion efficiency in the radio band initially increase slowly, and then decreases with $L_{tot}$.

\subsection{Constraining the Charge of a BH: Application for the GW200105 and GW200115
events} \label{subsec: charge of BH} Based on the Eq.(\ref{eq_L_e}) and Eq.(\ref{eq_LB}), the $L_{\rm tot}=L_{e}+L_{m}$ is dependent on the charge and mass of both the BH and NS. The charge of the NS ($Q_{NS}$) also depends on the parameters of the NS, e.g., period, the strength of magnetic field, and the radius, but we know little about those parameters. In order to test the magnitude of $Q_{NS}$, we fix the radius $R_{\rm NS}=12$ km, and adopt five groups of typical period ($P=$1 ms, 5 ms, 10 ms, 50 ms, and 100 ms) and strength of magnetic field ($B_p=10^{12}$ G, $10^{13}$ G, $10^{14}$ G, $10^{15}$ G, and $10^{16}$ G). Then we calculated the NS charge using various combinations of the above parameters.

Applying this analysis to the GW200105 and GW200115 events, we can estimate the parameters of the BH and NS (e.g., mass of the BH and NS, luminosity distance, mass ratio, and chirp mass) via the GW observations. On the other hand, by assuming that the observed optical transients or even upper limits in section 3 are the EM counterparts of the GW200105 and GW200115 events, then, we can calculate the luminosity ($L_{\rm obs}$) based on the luminosity distance\footnote{The luminosity distance is related to the parameters of cosmology, and it can be expressed as $D_L(z)=(1+z)\frac{c}{H_0}\int^z_0\frac{dz'}{\sqrt{\Omega_M(1+z')^3+\Omega_{\Lambda}}}$, where $z$ is the redshift.} and observed magnitude in the optical/IR band. Here, we calculate the $L_{\rm
obs}$ by adopting the maximum and minimum magnitudes in the optical/IR in table \ref{tab:EM candidates 200105} and \ref{tab:EM candidates 200115} as marked $L_{\rm obs,max}$ and $L_{\rm obs,min}$, respectively. The range of $L_{\rm obs}$ is $L_{\rm obs,min}<L_{\rm obs}<L_{\rm obs,max}$, and adopt $L_{\rm obs}$ corresponding $L_{\rm \nu}$. Based on the Fig.\ref{fig:L_eta}, one can roughly estimate the conversion efficiency in the optical/IR bands. Combining with Eqs.(\ref{eq_L_e}-\ref{eq_LB}), one can derive the luminosity of electric and magnetic dipole radiation at each orbital separation ($a$). Then, we take the sum of the luminosity at each orbital
separation as the injected energy of Poynting-flux-dominated outflow. Finally, by adopting Eqs.(\ref{eq_QNS}-\ref{eq_nua}), one can roughly estimate the charge on the BH for $L_{\rm \nu}=L_{\rm obs,min}$ and $L_{\rm \nu}=L_{\rm obs,max}$. The results are presented in Table \ref{tab:BH charge}. In this table, we define $\hat{q}_{BH,1}$ and $\hat{q}_{BH,2}$ in unit of $Q_{BH}/M_{BH}$ (the charge for unit mass) are derived by the magnitude of the brightest and dimmest EM candidates for GW 200105 and GW200115, respectively.

For GW200105, we find that the estimated maximal charge-to-mass ratio of BH is between
$9.73\times 10^{-8}$ and $1.12\times 10^{-6}{~\rm esu}$ with $B_{p}\lesssim 10^{16}$ G and $P>~1$ ms for the NS. Similarly, for GW200115, we also estimate the maximal charge-to-mass ratio of BH which ranges from $8.84\times 10^{-8}$ to $1.53\times 10^{-6}{~\rm esu}$ with $B_{p}\lesssim ~10^{16}$ G and $P>~1$ ms for the NS.

\section{Conclusion and discussion}\label{sec:Conclusion}
The physical properties of any BH can be described with its mass, angular momentum, and charge. The mass and angular momentum of the BH can be roughly measured based on currently observed data, but to infer the charge of a BH remains an open question. One proposal is that a charged BH and NS merger could be a potential approach to constrain the BH charge \citep{Zhang(2019)}. Recently, two GW events (GW200105 and GW200115) originating from the merger of a BH-NS system are confirmed to be detected by aLIGO and Virgo, and several optical source candidates are reported to possible associate with the two GW events, but not confirmed yet. In this paper, by assuming that the possible optical sources are associated with GW200105 and GW200115, we try to
estimate the BH charge via the observed upper limits of the EM radiation in the charged BH and NS system.

A charged BH merging with a NS can produce electric and magnetic dipole radiation. Then, the energy from the electric and magnetic dipole radiation injects into the outflow which is Poynting flux dominated, and the magnetic energy can convert into the kinetic energy of electrons by magnetic reconnection and turbulence to accelerate electrons. It can produce the observed EM radiation by assuming synchrotron radiation in the outflow. Within this scenario, we calculate the conversion efficiency of the electric dipole and magnetic dipole radiation within different energy bands, and constrain the BH charge for given NS physical parameters (e.g., period and surface magnetic field). The following interesting results are obtained.
\begin{itemize}
\item We find that the conversion efficiency in the X-ray band is much higher than that of the
    ultraviolet (UV), near-infrared, and radio bands, and it increases with $L_{tot}$ in
    the X-ray, UV, and near-infrared. However, the conversion efficiency in the radio band
    initially increases slowly, and then decreases with $L_{tot}$.
\item For GW200105, we find that the estimated maximal charge-to-mass ratio  (the charge for unit mass) of the BH is between about $9.73\times 10^{-8}$ and $1.12\times 10^{-6}$ esu with $B_{p}\lesssim ~10^{16}$ G and $P>~1$ ms for the NS. Similarly, for GW200115, we also estimate the maximal charge-to-mass ratio of the BH which ranges from $8.84\times 10^{-8}$ to $1.53\times 10^{-6}$ esu with $B_{p}\lesssim ~10^{16}$ G and $P>~1$ ms for the NS.
\end{itemize}

In fact, the radiation is only found in the leading order post-newtonian expansion, which for the electric dipole is even lower order (-1PN) than the GW quadrupole. In our calculations, we do not consider the contribution of EM force for modification of the orbit, because the EM force is much smaller than gravitational force in our analysis. However, the contribution of EM force is not negligible if EM and gravitational forces are comparable between each other \citep{Liu et al.(2020), Christiansen et al.(2021)}.

It has always been expected to be able to simultaneously observe GW and EM signals from BH-NS merger systems, and to search for the EM counterparts of the GW event from such mergers remains an interesting and hot topic in astrophysics. Reference \citep{Chen Dai(2021)} proposed that the NS charge can be transformed to the BH during inspiral phase of the BH-NS system. Within this scenario, the calculated BH charge in this work is underestimated, and it means that we do not consider the charge contributed by the magnetosphere of the NS within the BH-NS system. Within the scenario of charged BH-NS systems, the people have theoretically studied the EM and GW radiations in its
different evolutional phase through numerical Einstein-Maxwell equations. Reference \citep{Zilhao et al.(2012)} simulated charged-black-hole head-on collisions with the equal charge-to-mass ratio. They found that the ratio of energy carried between GW and EM radiations would decrease with the increasing of charge-to-mass ratio. Reference \citep{Bozzola Paschalidis(2021)} considered the quasi-circular inspiral and merger of charged BH-BH systems and derived the similar outcomes. These works stated that the EM force is negligible when the charge-to-mass ratio is small enough, and it is consistent
with our assumptions.

On the other hand, some researchers also try to constrain the charge of supermassive black holes that reside in the central of galaxies, e.g., Sgr A* and M87* \citep{Kocherlakota et al.(2021),Ghosh Afrin(2023)}. Reference \citep{Ghosh Afrin(2023)} considered Sgr A* as a Kerr–Newman black to constrain the supermassive black hole charge based on the Event Horizon Telescope (EHT) observation of the shadow. Similarly, References \citep{Karouzos(2018)} and \citep{Zajacek et al.(2018)} constrain the quantity of supermassive black hole charge which is less than $3.1 \times 10^8 C$ by using Chandra X-ray data.

\begin{figure*}
\centering
\includegraphics [angle=0,scale=0.6] {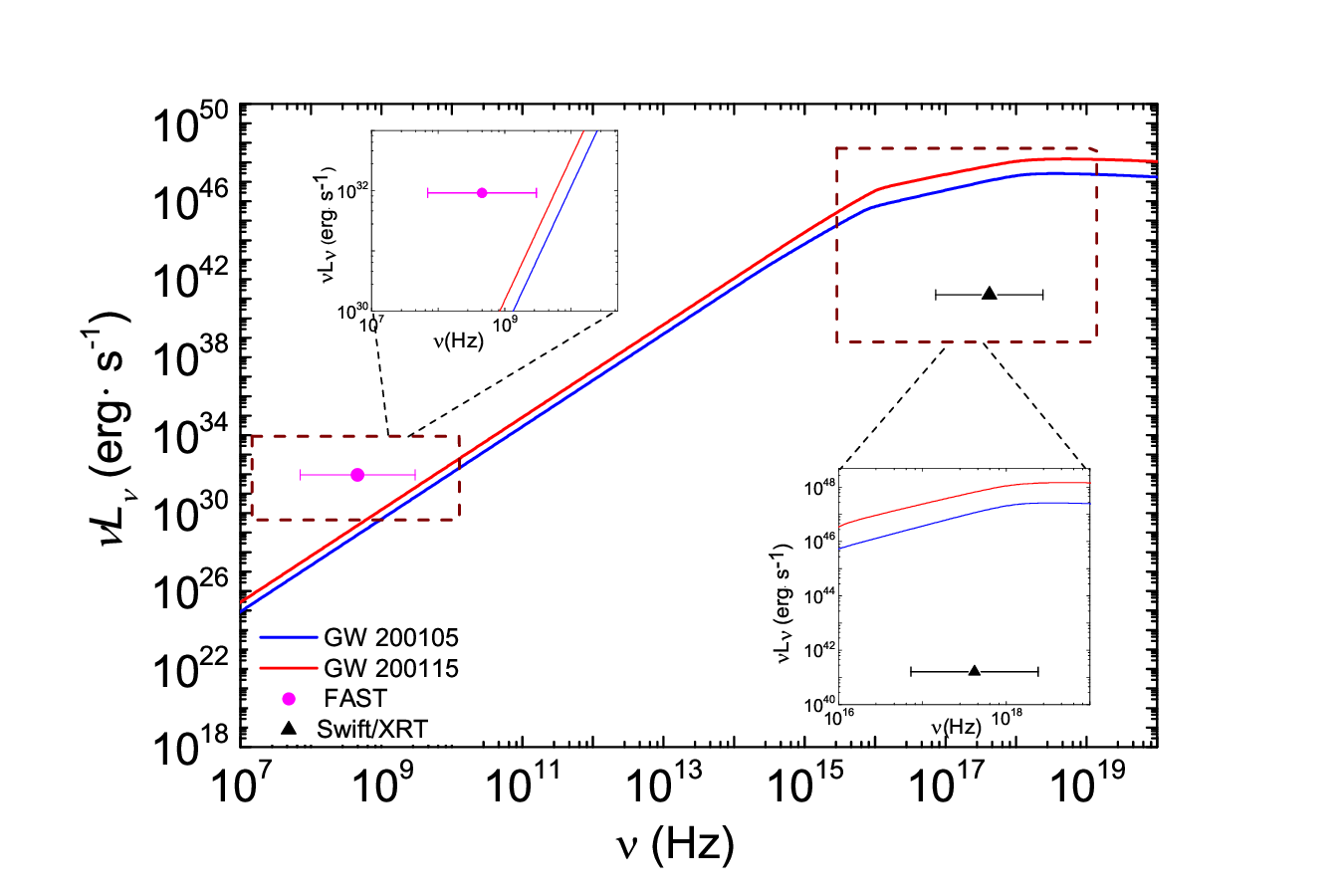}
\caption{Comparing the FAST and Swift/XRT sensitivity with the luminosity calculated by the
method mentioned in Sec.\ref{Sec: EM-counterpart}. We fix the NS parameters as $P=1$ ms and
$~B_{p}=10^{16}$ G for both GW200105 and
GW200115 events.}
\label{fig:FAST}
\end{figure*}

Moreover, by adopting the estimated parameters via possible optical counterparts for the GW200105 and GW200115 events, we calculate the possibility of detection in the radio and X-ray bands by comparing with the sensitivities of Five-hundred-meter Aperture Spherical radio Telescope (FAST) and Swift/XRT, respectively. We find that the FAST can not be triggered with fixed $P=1$ ms and $B_{p}=10^{16}$ G for the NS, but it is expected to detected by Swift/XRT. The reason may be the lower conversion efficiency of the dipole radiation in the radio band. The results are shown in Fig.\ref{fig:FAST}.

Owing to absence of a simultaneous observation of GW emission and any EM counterparts from a BH-NS merger system, to understand their properties remains an open question. However, it is necessary to study first in theory before discovering any GW emission associated with EM in the observations in such BH-NS systems. Due to the higher conversion efficiency of the dipole radiation in the X-ray band, we encourage the space based X-ray telescopes to follow up observations of GW events from BH-NS merger systems. \citet{Troja et al.(2010)} stated that there would be a precursor emission before
merger of binary neutron star. If the dipole radiations really exist during the inspiral and merger of charged binary compact star, the precursor emission would be another way to test our research and probe the properties of charged compact star systems. Moreover, the next generation GW detectors, e.g., Cosmic Explorer \citep{Reitze et al.(2019)} and the Einstein Telescope \citep{Punturo et al.(2010a), Punturo et al.(2010b)}, are also expected to be more sensitive and the higher sky localization ability will provide guidance for the follow-up EM observations.

One needs to note that the fundamental of this work is the existing BH-NS events. The several candidates (GW 190426, GW 190814, GW 190917, GW 191219, and GW 200210) as the BH-NS events remain in debate due to the high mass of secondary \citep{The LIGO Scientific Collaboration et al.(2021b)}. Also, GW 200105 is also debating that whether it is a BH-NS event or not due to the high likelihood of detector noise \citep{The LIGO Scientific Collaboration et al.(2021b)}. The LIGO pipeline considers anything below 3 solar mass as a neutron star, while the maximum mass of neutron stars is nearly certainly lower than that. So that, we expect that more confirmed BH-NS events can be detected by LVK in the future.

\begin{acknowledgments}
We acknowledge the use of the public data from the GRB Coordinates Network(GCN) archive. This work is supported by the National Natural Science Foundation of China (grant Nos.11922301, and 12133003), the Guangxi Science Foundation (grant Nos. 2023GXNSFDA026007, and 2017GXNSFFA198008), and the Program of Bagui Scholars Program (LHJ).
\end{acknowledgments}

	\nocite{*}
	
	\providecommand{\noopsort}[1]{}\providecommand{\singleletter}[1]{#1}%
	%
	
\clearpage

\begin{table}[h]

\centering
\caption{Properties of BH-NS merger events and their candidate objects}
\label{tab:GW properties}
\begin{tabular}{llllll}
\hline
GW event & $M_{NS}(M_{\odot})$    & $M_{BH}(M_{\odot})$  & redshift                  & mass
ratio ($q$) &
chirp mass($M_{\odot}$)             \\ \hline
GW200105 & $1.9^{+0.2}_{-0.2}$    & $8.9^{+1.1}_{-1.3}$  & $0.06^{+0.02}_{-0.02}$    &
$0.21_{-0.04}^{+0.03}$
& $3.41_{-0.07}^{+0.08}$ \\
GW200115 & $1.4^{+0.6}_{-0.2}$    & $5.9^{+1.4}_{-2.1}$  & $0.07^{+0.03}_{-0.02}$    &
$0.24_{-0.09}^{+0.12}$
& $2.42_{-0.07}^{+0.05}$ \\ \hline
GW190426 & $1.5^{+0.8}_{-0.5}$    & $5.7^{+3.9}_{-2.3}$  & $0.08^{+0.04}_{-0.03}$    &
$0.26_{-0.14}^{+0.23}$
& $2.41_{-0.08}^{0.08}$  \\
GW190814 & $2.59^{+0.08}_{-0.09}$ & $23.2^{+1.1}_{-1.0}$ & $0.050^{+0.009}_{-0.010}$ &
$0.11_{-0.01}^{+0.01}$
& $6.09_{-0.06}^{+0.06}$ \\
GW190917 & $2.1^{+1.5}_{-0.5}$    & $9.3^{+3.4}_{-4.4}$  & $0.15^{+0.06}_{-0.06}$    &
$0.23_{-0.12}^{+0.18}$
& $3.7_{-0.2}^{+0.2}$    \\
GW191219 & $1.17^{+0.06}_{-0.05}$ & $31.6^{+1.8}_{-2.5}$ & $0.11^{+0.04}_{-0.03}$    &
$0.037_{-0.003}^{+0.003}$    & $4.33_{-0.15}^{+0.10}$ \\
GW200210 & $2.79^{+0.54}_{-0.48}$ & $24.5^{+8.9}_{-5.3}$ & $0.19^{+0.07}_{-0.06}$    &
$0.11_{-0.03}^{+0.05}$
& $6.56_{-0.38}^{+0.34}$ \\ \hline
\end{tabular}
\end{table}

\begin{table*}[]
 \centering \caption{EM candidates for GW200105}
 \label{tab:EM candidates 200105}
\begin{threeparttable}
\resizebox{\textwidth}{60mm}{
\begin{tabular}{c|c|cccc}

\hline \multicolumn{1}{l|}{Detector} & object name            & RA             & Dec
&
filter & mag(AB) \\ \hline
\multirow{23}{*}{ZTF\tnote{a}}            & ZTF20aaervoa/AT2020pp & 15h 02m 38.38s & +16d
28m 21.5s  & r  &
20.08               \\
                              & ZTF20aaertpj/AT2020pv  & 14h 27m 52.03s & +33d 34m 09.7s  &
                              r      & 19.55
                              \\
                              & ZTF20aaervyn/AT2020pq  & 15h 01m 27.45s & +20d 37m 23.5s  &
                              r      & 20.49
                              \\
                              & ZTF20aaerqbx/AT2020ps  & 15h 49m 26.29s & +40d 49m 55.0s  &
                              r      & 19.70
                              \\
                              & ZTF20aaerxsd/AT2020py  & 14h 00m 54.27s & +45d 28m 21.8s  &
                              r      & 19.97
                              \\
                              & ZTF20aaexpwt/AT2020adi & 06h 26m 01.30s & +11d 33m 38.85s &
                              r      & 19.80
                              \\
                              & ZTF20aaertil/AT2020pu  & 14h 52m 25.86s & +31d 01m 18.50s &
                              g      & 19.70
                              \\
                              & ZTF20aaevbzl/AT2020adf & 13h 26m 41.27s & +30d 52m 30.68s &
                              g      & 20.30
                              \\
                              & ZTF20aagiipi/AT2020adl & 15h 33m 24.61s & +42d 02m 36.70s &
                              r      & 20.10
                              \\
                              & ZTF20aagijez/AT2020adm & 15h 04m 13.18s & +27d 29m 04.19s &
                              r      & 19.90
                              \\
                              & ZTF20aagiiik/AT2020abl & 16h 19m 09.94s & +53d 45m 38.26s &
                              r      & 20.30
                              \\
                              & ZTF20aafexle/AT2020adn & 04h 20m 31.22s & -09d 30m 28.20s &
                              r      & 19.90
                              \\
                              & ZTF20aaflndh/AT2020xz  & 01h 22m 38.14s & -06d 49m 34.31s &
                              g      & 19.20
                              \\
                              & ZTF20aafduvt/AT2020ado & 03h 36m 28.54s & -07d 49m 34.52s &
                              g      & 20.10
                              \\
                              & ZTF20aafefxe/AT2020adt & 07h 47m 24.18s & +14d 42m 23.82s &
                              r      & 20.90
                              \\
                              & ZTF20aafkkoy/AT2020adp & 13h 40m 59.99s & +40d 48m 49.03s &
                              r      & 19.30
                              \\
                              & ZTF20aafaoki/AT2020adq & 05h 13m 13.80s & +05d 09m 56.55s &
                              r      & 19.60
                              \\
                              & ZTF20aafujqk/AT2020adg & 17h 57m 00.41s & +10d 32m 20.31s &
                              r      & 18.60
                              \\
                              & ZTF20aafksha/AT2020adr & 13h 43m 54.45s & +38d 25m 13.82s &
                              g      & 20.00
                              \\
                              & ZTF20aagjemb/AT2020adh & 14h 51m 25.85s & +45d 20m 41.18s &
                              r      & 20.30
                              \\
                              & ZTF20aafdxkf/AT2020ads & 03h 42m 07.08s & -03d 11m 39.05s &
                              r      & 20.20
                              \\
                              & ZTF20aafukgx/AT2020adj & 18h 23m 21.44s & +17d 49m 31.65s &
                              r      & 19.20
                              \\
                              & ZTF20aafanxk/AT2020adk & 05h 35m 36.04s & +11d 46m 15.31s &
                              r      & 19.40
                              \\ \hline
\multicolumn{1}{l|}{GTC\tnote{b}}         & AT2020oo/Gaia20add    & 17h 19m 6.18s & +25d 27m
20.23s
& r' & 17.56               \\ \hline \multicolumn{1}{c|}{MASTER-SAAO\tnote{c}} &         /
& 04h 16m 26.77s & -54d 52m 08.5s  & V  & \textgreater{}16.00 \\ \hline
\end{tabular}}
\begin{tablenotes}
        \footnotesize
        \item[a] The observational data for this detector are collected from \citet{Stein et al.(2020)} and
            \citet{Ahumada(2020)}.
        \item[b] The observational data for this detector are collected from \citet{Hu et al.(2020)}.
        \item[c] This upper limit of the detector is collected from \citet{Lipunov et al.(2020)}.
\end{tablenotes}
\end{threeparttable}
\end{table*}

\begin{table*}[]
\centering \caption{EM candidates for GW200115} \label{tab:EM candidates 200115}
\begin{threeparttable}
\begin{tabular}{c|c|cccc}
\hline
Detector                     & object name           & RA             & Dec            &
filter & mag(AB) \\
\hline
\multirow{3}{*}{ZTF\tnote{a}}         & ZTF20aafqpum/AT2020yo & 03h 06m 06.50s    & +13d 54m
48.4s    & g
& 20.80   \\
                             & ZTF20aafqulk/AT2020yp & 03h 39m 45.43s    & +27d 44m 05.4s
                             & g      &
                             21.10
                             \\
                             & ZTF20aafqvyc/AT2020yq & 03h 47m 58.21s    & +38d 26m 31.8s
                             & g      &
                             20.90
                             \\ \hline
Swift/UVOT\tnote{b}                   & S200115j\_X136        & 02h 40m 12.19s & -02d 33m
45.4s & u      &
16.54   \\ \hline
\multirow{8}{*}{Pan-STARRS2\tnote{c}} & PS20ev/AT2020ait      & 02h 36m 56.51s    & -02d 12m
03.5s    &
w\tnote{*}      & 20.23   \\
                             & PS20fs/AT2020ajy      & 02h 53m 21.00s    & +03d 23m 17.4s
                             & w      &
                             21.46
                             \\
                             & PS20fo/AT2020ajw      & 02h 25m 53.75s    & -03d 13m 04.6s
                             & w      &
                             19.90
                             \\
                             & PS20fr/AT2020ajx      & 02h 25m 23.51s    & -07d 50m 15.5s
                             & w      &
                             20.42
                             \\
                             & PS20fu/AT2020ajz      & 02h 49m 02.17s    & +04d 59m 11.6s
                             & w      &
                             21.16
                             \\
                             & PS20fv/AT2020aka      & 02h 37m 24.50s    & -01d 09m 20.1s
                             & w      &
                             19.91
                             \\
                             & PS20fw/AT2020akb      & 03h 10m 05.41s    & +05d 59m 16.4s
                             & w      &
                             21.05
                             \\
                             & PS20f/AT2020akc      & 02h 52m 35.05s    & +07d 15m 11.9s
                             & w      & 20.89
                             \\ \hline
\end{tabular}
\begin{tablenotes}
        \footnotesize
        \item[*] The ``pivot'' wavelength of the w-band is $0.608\mu m$ \citep{Tonry et al.(2012)}.
        \item[a] The relevant data are in \citet{Anand et al.(2020)}.
        \item[b] Swift/UVOT observational data are from \citet{Oates Swift Team(2020)}.
        \item[c] The relevant data are in \citet{Srivastav Smartt(2020)}.
\end{tablenotes}
\end{threeparttable}

\end{table*}

\begin{table*}[]
\centering \caption{BH charge for GW200105 and GW200115 with different NS $B_p$ and $P$.}
\label{tab:BH charge}
\begin{threeparttable}
\begin{tabular}{ccccccc}
\hline
 &
   &
  \multicolumn{5}{c}{GW200105} \\ \hline
\multicolumn{1}{c|}{\diagbox{$B_p$ (G)}{$\hat{q}_{BH}(esu)$\tnote{*}}{P~(ms)}} &
   &
  \multicolumn{1}{c|}{1} &
  \multicolumn{1}{c|}{5} &
  \multicolumn{1}{c|}{10} &
  \multicolumn{1}{c|}{50} &
  100 \\ \hline
\multicolumn{1}{c|}{\multirow{2}{*}{$10^{12}$}} &
  \multicolumn{1}{c|}{$\hat{q}_{BH,1}$} &
  \multicolumn{1}{c|}{$4.86\times 10^{-7}$} &
  \multicolumn{1}{c|}{$4.86\times 10^{-7}$} &
  \multicolumn{1}{c|}{$4.86\times 10^{-7}$} &
  \multicolumn{1}{c|}{$4.86\times 10^{-7}$} &
  $4.86\times 10^{-7}$ \\
\multicolumn{1}{c|}{} &
  \multicolumn{1}{c|}{$\hat{q}_{BH,2}$} &
  \multicolumn{1}{c|}{$9.74\times 10^{-8}$} &
  \multicolumn{1}{c|}{$9.73\times 10^{-8}$} &
  \multicolumn{1}{c|}{$9.73\times 10^{-8}$} &
  \multicolumn{1}{c|}{$9.73\times 10^{-8}$} &
  $9.73\times 10^{-8}$ \\ \hline
\multicolumn{1}{c|}{\multirow{2}{*}{$10^{13}$}} &
  \multicolumn{1}{c|}{$\hat{q}_{BH,1}$} &
  \multicolumn{1}{c|}{$4.87\times 10^{-7}$} &
  \multicolumn{1}{c|}{$4.86\times 10^{-7}$} &
  \multicolumn{1}{c|}{$4.86\times 10^{-7}$} &
  \multicolumn{1}{c|}{$4.86\times 10^{-7}$} &
  $4.86\times 10^{-7}$ \\
\multicolumn{1}{c|}{} &
  \multicolumn{1}{c|}{$\hat{q}_{BH,2}$} &
  \multicolumn{1}{c|}{$9.79\times 10^{-8}$} &
  \multicolumn{1}{c|}{$9.74\times 10^{-8}$} &
  \multicolumn{1}{c|}{$9.74\times 10^{-8}$} &
  \multicolumn{1}{c|}{$9.74\times 10^{-8}$} &
  $9.73\times 10^{-8}$ \\ \hline
\multicolumn{1}{c|}{\multirow{2}{*}{$10^{14}$}} &
  \multicolumn{1}{c|}{$\hat{q}_{BH,1}$} &
  \multicolumn{1}{c|}{$4.93\times 10^{-7}$} &
  \multicolumn{1}{c|}{$4.87\times 10^{-7}$} &
  \multicolumn{1}{c|}{$4.87\times 10^{-7}$} &
  \multicolumn{1}{c|}{$4.86\times 10^{-7}$} &
  $4.86\times 10^{-7}$ \\
\multicolumn{1}{c|}{} &
  \multicolumn{1}{c|}{$\hat{q}_{BH,2}$} &
  \multicolumn{1}{c|}{$1.04\times 10^{-7}$} &
  \multicolumn{1}{c|}{$9.86\times 10^{-8}$} &
  \multicolumn{1}{c|}{$9.79\times 10^{-8}$} &
  \multicolumn{1}{c|}{$9.74\times 10^{-8}$} &
  $9.74\times 10^{-8}$ \\ \hline
\multicolumn{1}{c|}{\multirow{2}{*}{$10^{15}$}} &
  \multicolumn{1}{c|}{$\hat{q}_{BH,1}$} &
  \multicolumn{1}{c|}{$5.50\times 10^{-7}$} &
  \multicolumn{1}{c|}{$4.99\times 10^{-7}$} &
  \multicolumn{1}{c|}{$4.93\times 10^{-7}$} &
  \multicolumn{1}{c|}{$4.87\times 10^{-7}$} &
  $4.87\times 10^{-7}$ \\
\multicolumn{1}{c|}{} &
  \multicolumn{1}{c|}{$\hat{q}_{BH,2}$} &
  \multicolumn{1}{c|}{$1.61\times 10^{-7}$} &
  \multicolumn{1}{c|}{$1.10\times 10^{-7}$} &
  \multicolumn{1}{c|}{$1.04\times 10^{-7}$} &
  \multicolumn{1}{c|}{$9.86\times 10^{-8}$} &
  $9.79\times 10^{-8}$ \\ \hline
\multicolumn{1}{c|}{\multirow{2}{*}{$10^{16}$}} &
  \multicolumn{1}{c|}{$\hat{q}_{BH,1}$} &
  \multicolumn{1}{c|}{$1.12\times 10^{-6}$} &
  \multicolumn{1}{c|}{$6.13\times 10^{-7}$} &
  \multicolumn{1}{c|}{$5.50\times 10^{-7}$} &
  \multicolumn{1}{c|}{$4.99\times 10^{-7}$} &
  $4.93\times 10^{-7}$ \\
\multicolumn{1}{c|}{} &
  \multicolumn{1}{c|}{$\hat{q}_{BH,2}$} &
  \multicolumn{1}{c|}{$5.38\times 10^{-7}$} &
  \multicolumn{1}{c|}{$2.24\times 10^{-7}$} &
  \multicolumn{1}{c|}{$1.61\times 10^{-7}$} &
  \multicolumn{1}{c|}{$1.10\times 10^{-7}$} &
  $1.04\times 10^{-7}$ \\ \hline
 &
   &
  \multicolumn{5}{c}{GW200115} \\ \hline
\multicolumn{1}{c|}{\multirow{2}{*}{$10^{12}$}} &
  \multicolumn{1}{c|}{$\hat{q}_{BH,1}$} &
  \multicolumn{1}{c|}{$6.69\times 10^{-7}$} &
  \multicolumn{1}{c|}{$6.69\times 10^{-7}$} &
  \multicolumn{1}{c|}{$6.69\times 10^{-7}$} &
  \multicolumn{1}{c|}{$6.69\times 10^{-7}$} &
  $6.69\times 10^{-7}$ \\
\multicolumn{1}{c|}{} &
  \multicolumn{1}{c|}{$\hat{q}_{BH,2}$} &
  \multicolumn{1}{c|}{$8.85\times 10^{-8}$} &
  \multicolumn{1}{c|}{$8.84\times 10^{-8}$} &
  \multicolumn{1}{c|}{$8.84\times 10^{-8}$} &
  \multicolumn{1}{c|}{$8.84\times 10^{-8}$} &
  $8.84\times 10^{-8}$ \\ \hline
\multicolumn{1}{c|}{\multirow{2}{*}{$10^{13}$}} &
  \multicolumn{1}{c|}{$\hat{q}_{BH,1}$} &
  \multicolumn{1}{c|}{$6.70\times 10^{-7}$} &
  \multicolumn{1}{c|}{$6.69\times 10^{-7}$} &
  \multicolumn{1}{c|}{$6.69\times 10^{-7}$} &
  \multicolumn{1}{c|}{$6.69\times 10^{-7}$} &
  $6.69\times 10^{-7}$ \\
\multicolumn{1}{c|}{} &
  \multicolumn{1}{c|}{$\hat{q}_{BH,2}$} &
  \multicolumn{1}{c|}{$8.93\times 10^{-8}$} &
  \multicolumn{1}{c|}{$8.86\times 10^{-8}$} &
  \multicolumn{1}{c|}{$8.85\times 10^{-8}$} &
  \multicolumn{1}{c|}{$8.84\times 10^{-8}$} &
  $8.84\times 10^{-8}$ \\ \hline
\multicolumn{1}{c|}{\multirow{2}{*}{$10^{14}$}} &
  \multicolumn{1}{c|}{$\hat{q}_{BH,1}$} &
  \multicolumn{1}{c|}{$6.78\times 10^{-7}$} &
  \multicolumn{1}{c|}{$6.71\times 10^{-7}$} &
  \multicolumn{1}{c|}{$6.70\times 10^{-7}$} &
  \multicolumn{1}{c|}{$6.69\times 10^{-7}$} &
  $6.69\times 10^{-7}$ \\
\multicolumn{1}{c|}{} &
  \multicolumn{1}{c|}{$\hat{q}_{BH,2}$} &
  \multicolumn{1}{c|}{$9.70\times 10^{-8}$} &
  \multicolumn{1}{c|}{$9.01\times 10^{-8}$} &
  \multicolumn{1}{c|}{$8.93\times 10^{-8}$} &
  \multicolumn{1}{c|}{$8.86\times 10^{-8}$} &
  $8.85\times 10^{-8}$ \\ \hline
\multicolumn{1}{c|}{\multirow{2}{*}{$10^{15}$}} &
  \multicolumn{1}{c|}{$\hat{q}_{BH,1}$} &
  \multicolumn{1}{c|}{$7.55\times 10^{-7}$} &
  \multicolumn{1}{c|}{$6.86\times 10^{-7}$} &
  \multicolumn{1}{c|}{$6.78\times 10^{-7}$} &
  \multicolumn{1}{c|}{$6.71\times 10^{-7}$} &
  $6.70\times 10^{-7}$ \\
\multicolumn{1}{c|}{} &
  \multicolumn{1}{c|}{$\hat{q}_{BH,2}$} &
  \multicolumn{1}{c|}{$1.75\times 10^{-7}$} &
  \multicolumn{1}{c|}{$1.06\times 10^{-7}$} &
  \multicolumn{1}{c|}{$9.70\times 10^{-8}$} &
  \multicolumn{1}{c|}{$9.01\times 10^{-8}$} &
  $8.93\times 10^{-8}$ \\ \hline
\multicolumn{1}{c|}{\multirow{2}{*}{$10^{16}$}} &
  \multicolumn{1}{c|}{$\hat{q}_{BH,1}$} &
  \multicolumn{1}{c|}{$1.53\times 10^{-6}$} &
  \multicolumn{1}{c|}{$8.41\times 10^{-7}$} &
  \multicolumn{1}{c|}{$7.55\times 10^{-7}$} &
  \multicolumn{1}{c|}{$6.86\times 10^{-7}$} &
  $6.78\times 10^{-7}$ \\
\multicolumn{1}{c|}{} &
  \multicolumn{1}{c|}{$\hat{q}_{BH,2}$} &
  \multicolumn{1}{c|}{$9.50\times 10^{-7}$} &
  \multicolumn{1}{c|}{$2.61\times 10^{-7}$} &
  \multicolumn{1}{c|}{$1.75\times 10^{-7}$} &
  \multicolumn{1}{c|}{$1.06\times 10^{-7}$} &
  $9.70\times 10^{-8}$ \\ \hline
\end{tabular}
\begin{tablenotes}
        \footnotesize
\item[*] Derived the BH charge for given $B_p$ and $P$ of NS. $\hat{q}_{BH,1}$ and
    $\hat{q}_{BH,2}$ \textbf{in unit of $Q_{BH}/M_{BH}$ } (the charge for unit mass) are derived by the magnitude of the brightest and dimmest EM candidates for GW 200105
    and GW200115, respectively.
\end{tablenotes}
\end{threeparttable}
\end{table*}

	
\end{document}